# An On-Chip Quad-Wavelength Pyroelectric Sensor for Spectroscopic Infrared Sensing

Thang Duy Dao[*,1], Satoshi Ishii[1], Anh Tung Doan[1,2], Yoshiki Wada[3], Akihiko Ohi[4], Toshihide Nabatame[4], Tadaaki Nagao[*,1,2]

[1]*International Center for Materials Nanoarchitectonics (MANA), National Institute for Materials Science (NIMS), 1-1 Namiki, Tsukuba, Ibaraki 305-0044, Japan.*
[2]*Department of Condensed Matter Physics, Graduate School of Science, Hokkaido University, Kita 10, Nisi 8, Kita-ku, Sapporo 060-0810, Japan.*
[3]*Research Center for Functional Materials, National Institute for Materials Science (NIMS), 1-1 Namiki, Tsukuba, Ibaraki 305-0044, Japan.*
[4]*Nanotechnology Innovation Station, National Institute for Materials Science (NIMS), 1-1 Namiki, Tsukuba, Ibaraki 305-0044, Japan.*

[*]*Corresponding Authors*
*Thang Duy Dao: Dao.duythang@nims.go.jp; katsiusa@gmail.com*
*Tadaaki Nagao: Nagao.Tadaaki@nims.go.jp*

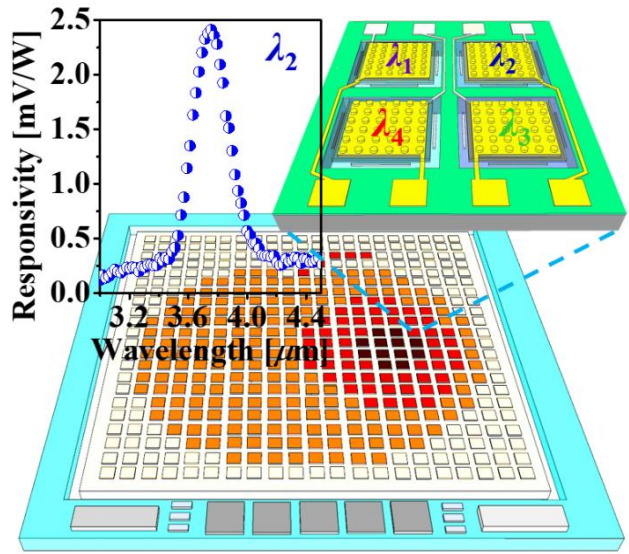

**Abstract.** Merging photonic structures and optoelectronic sensors into a single chip may yield a sensor-on-chip spectroscopic device that can measure the spectrum of matters. In this work, we propose and realize an on-chip concurrent multi-wavelength infrared (IR) sensor which consist of a set of narrowband wavelength-selective plasmonic perfect absorbers combined with pyroelectric sensors, where the response of each pyroelectric sensor is boosted only at the resonance of the nanostructured absorber. The proposed absorber, which is based on the Wood's anomaly absorption from a two-dimensional plasmonic square lattice, shows a narrowband polarization-independent resonance (quality factor – $Q$ of 73) with a nearly perfect absorptivity as high as 0.99 at normal incidence. The fabricated quad-wavelength IR sensors exhibit four different narrowband spectral responses at normal incidence following the pre-designed resonances in the mid-wavelength infrared region that corresponds to the atmospheric window. The device can be applied for practical spectroscopic applications such as non-dispersive IR sensors, IR chemical imaging devices, pyrometers, and spectroscopic thermography imaging.

1. Introduction

Infrared spectroscopy is one of the most powerful spectroscopic technique to identify chemical species by analyzing their vibrational absorptions, as well as to characterize optical properties of narrow band-gap semiconductors, quantum well emitters, and thermal absorbers. The performance of an IR spectrometer, which includes the resolution, sensitivity and measurement time, relies on the massive spectroscopic modules such as prisms, diffraction gratings, interferometers, mechanical scanners and goniometers, as well as on their detection modules. Recent advances in photonics and nanofabrication techniques can lead to other types of IR spectroscopic devices in which their dispersive elements and IR detector are integrated into a compact microdevice for portable yet accurate spectroscopic applications; such as non-dispersive infrared (NDIR) sensors, dual-wavelength pyrometers, as well as chemical IR imaging integrated with thermography devices. The most important requirement of such devices is the dispersive element working at normal incidence, which must be simple, and has a narrowband width with a clean background in a broad spectral range. Thence, the interference filter is one of the most common choices. However, adding such macroscopic interference filters before IR sensors limits the number of the resonant wavelength for precise and multifunctional IR spectroscopic devices. Another approach that has been advanced as a possible solution to the requirement of this handheld concurrent multi-wavelength IR sensor, is to employ novel photonic crystals[1–4] and plasmonic structures[5–10], wherein the field confinement and resonant bandwidth as well as the resonant tunability can be controlled more easily in micron-scale devices. Nevertheless, the most photonic and plasmonic structures are rather complex structures and have been applied only for individual single-wavelength IR sensors. Moreover, their bandwidths are far wider than those of conventional spectrometers that possess resolutions narrower than the widths of molecular vibrations. These are the major impediments for realizing miniature sensor-on-chip spectroscopic devices, and thus, simple small-size spectroscopic modules with ultra-narrowband detection and broad tunability in the IR spectral range is strongly required.

In the recent years, resonant plasmonic metamaterial absorbers have attracted a great interest in the field of photonics due to their versatile ability in achieving near-unity absorption and in controlling the resonant bandwidth as well as tunability.[11] They have shown impressive practical applications involving solar energy harvesting,[12–14] thermal photovoltaics,[15–17] thermal emission,[18–21] radiative cooling,[22–24] NDIR spectroscopy,[25,26] amplifying signals in IR spectroscopy,[27,28] as well as IR sensors[29–34]. Particularly, when an electromagnetic field is absorbed by a perfect absorber, the absorbed energy is eventually converted into heat through Joule heating that follows Poynting's theorem:[35,36]

$$\frac{1}{2}\frac{\partial}{\partial t}\left(\varepsilon_0 \vec{E}\cdot\vec{E} + \mu_0 \vec{H}\cdot\vec{H}\right) + \nabla\cdot\left(\vec{E}\times\vec{H}\right) + \vec{J}\cdot\vec{E} = 0, \qquad (1)$$

where $\varepsilon_0$ and $\mu_0$ are the electric constant and the magnetic constant in vacuum, respectively. $\vec{E}$ is the electric field, $\vec{H}$ is the magnetic field, and $\vec{J}$ is the current density corresponding to the motion of charge. The first term $\frac{1}{2}\frac{\partial}{\partial t}\left(\varepsilon_0\vec{E}\cdot\vec{E}+\mu_0\vec{H}\cdot\vec{H}\right)$ is the time-averaged dissipative energy density. The term $\vec{J}\cdot\vec{E}$ is so called Joule heating that represents the power absorbed per unit volume. At the resonance of a perfect absorber, the term $\nabla\cdot\left(\vec{E}\times\vec{H}\right)$, which represents energy flux leaving the absorber, is almost zero, thus the dissipative energy density is simply calculated by, $\frac{1}{2}\varepsilon_0\omega\,\mathrm{Im}\,\varepsilon(\omega)|\vec{E}|^2$, where $\omega$ is the angular frequency, $\varepsilon(\omega)$ is the dielectric function of the non-magnetic dispersive and absorptive (dielectric) medium. This indicates that the absorbed energy of a perfect absorber irradiated at the resonance is proportional to the working frequency and the imaginary part of the dielectric function of the medium.[37–39] Thus, a perfect absorber at the resonance can be an efficient light-to-heat transducer. By integrating individual single-wavelength resonant perfect absorbers with thermal detectors into a single chip, the multi-wavelength IR devices can be feasible.

In this work, we first discuss the possible absorber structures that are compatible for on-chip multi-wavelength thermal sensors. Then, we propose a conceptual design of sensor-on-chip infrared spectroscopic devices utilizing Wood's anomaly absorption from two dimensional (2D) periodic metallic arrays, which are directly mounted on individual pyroelectric ZnO transducer for efficient light-to-heat conversion and efficient heat transfer. The 2D plasmonic absorber exhibits an ultra-narrowband polarization-independent resonance at the normal incidence in the mid-wavelength infrared (MWIR) region with a high $Q$-factor of 73, and with a nearly perfect absorptivity as high as 0.99. As a proof of concept of the spectroscopic microdevices for next generation MWIR sensors, we fabricated a set of on-chip, membrane-supported quad-wavelength infrared sensors on a 3-inch Si wafer. The fabricated quad-wavelength infrared sensors exhibited sharp spectral responses at the wavelengths exactly matching to the resonance wavelengths at the normal incidence. We also provide a detailed fabrication procedure of the device, which can be also applicable for scaling down the device to sub-hundred micrometer scale, making the proposed devices suitable for miniature spectroscopic sensors.

**2. Design Strategy of Ultra-Narrowband Perfect Absorbers Aiming for Wavelength-Selective Thermal Sensors**

In the beginning, we examine four different perfect absorber configurations that are compatible with thermal sensors for sensor-on-chip MWIR spectroscopic devices. The requirements to achieve a wavelength-selective absorber are the confined resonator and the intrinsic loss of the medium. In this concern, resonators consisted of a lossless dielectric and a low-loss metal are the most common structures for perfect absorbers.[11,18–20] Figure 1 indicates four different perfect absorbers that can be combined with thermal sensors such as pyroelectric, bolometric or thermoelectric materials for on-chip IR spectroscopic devices. As seen in Figure 1a, a metal-insulator-metal absorber with a plasmonic dipole resonator array on the top shows a perfect resonant absorption peak with a $Q$-factor of 10. In this configuration, the radiation is absorbed in both the top resonant antenna and the bottom metal mirror. Due to the low-thermal conductivity of the insulator, the heat induced at the top resonant antenna may not efficiently transfer to the thermal sensor. Figure 1b shows another perfect absorber configuration utilizing the nearfield confinement in a tiny-gap array that exhibits the similar $Q$-factor and absorptivity. This design can improve the conductive heat transfer from the absorber to the thermal sensor. However, this structure requires the gaps to be a few tens of nanometers in width and 500 nm in height, which is quite challenging to fabricate. It should be noted that, the resonant bandwidths of these absorbers (Figure 1a and b) are much broader compared to that of typical molecular vibrations, which severely limits the application of these spectroscopic sensors.

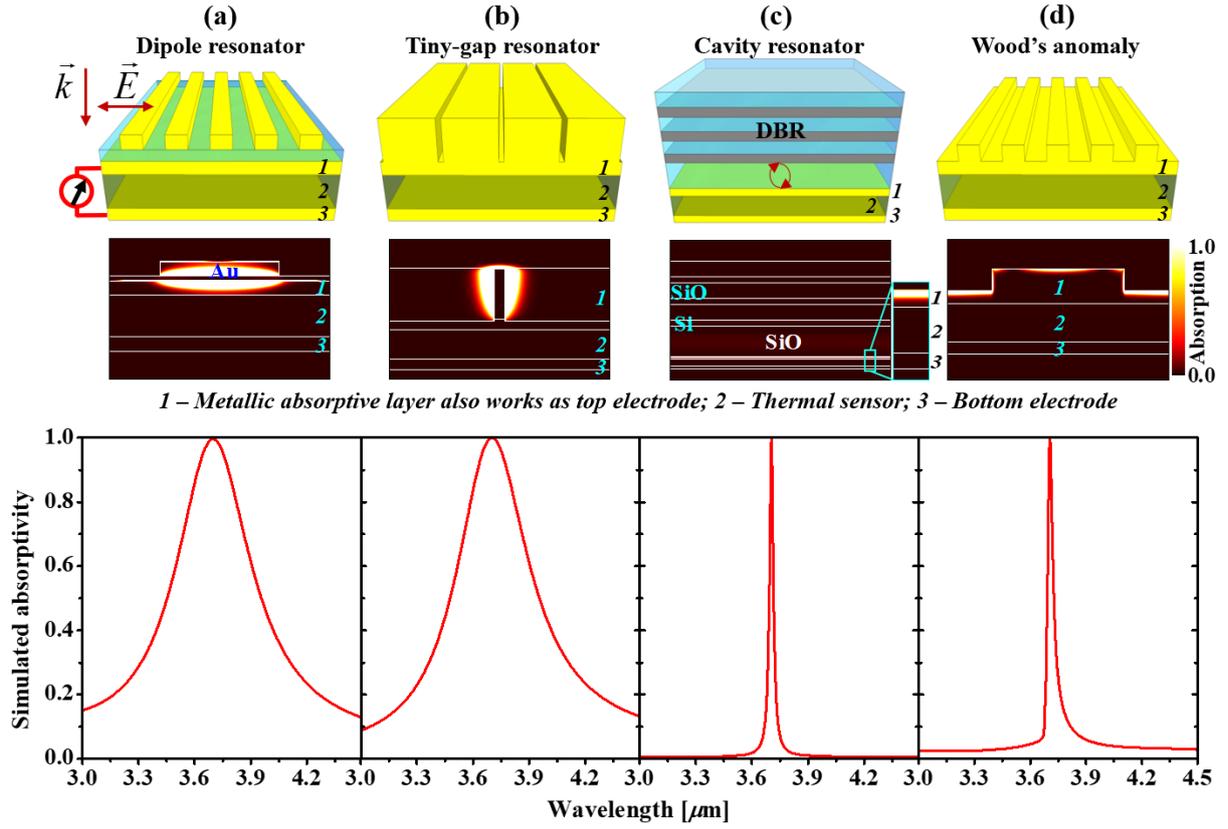

**Figure 1.** Comparison of four different perfect absorbers for IR wavelength-selective thermal sensors which have resonances at around 3.7 $\mu$m. From top to bottom: device structures, simulated absorption maps, and absorptivity spectra of the four absorbers. (a) Metal dipole antenna-insulator-metal absorber. (b) Tiny-gap plasmonic resonator array exhibiting unity absorption peak with *Q*-factors of about 10. (c) Asymmetric cavity absorber with the top reflector made of a distributed Bragg reflector (DBR). (d) Wood's anomaly (grating-coupled SPP) exhibiting much narrower unity absorption peak with *Q*-factor as high as ~150.

In contrast, absorbers incorporating periodic structures can significantly narrow the resonance bandwidth. For examples, Figure 1c and d show two different periodic structures that exhibit ultra-narrowband resonance with *Q*-factors as high as 150, which are comparable to the most molecular vibrations in solid state. Since the sharp resonance of the cavity absorber shown in Figure 1c is embodied by the layered photonic structure, it is not straightforward to fabricate cavities having different resonance wavelengths on a single chip. Keeping this limitation in mind, in Figure 1d we consider Wood's anomaly absorption in a 1D metallic grating structure which is realized by the surface-wave resonance condition between the surface plasmon polariton (SPP) and 1D grating.[40–44] The heat induced in the metallic array directly transfers to the thermal sensor module without being blocked by low thermal-conductivity object. The perfect absorption is realized by designing the depth and shape of the unit cell structure, and its resonance wavelength and directivity can be flexibly tuned by its periodicity. These features make the Wood's anomaly perfect absorber advantageous compared to the other three absorbers in the on-chip multi-wavelength IR sensor application. It should be noted that the three 1D periodic absorbers shown in Figure 1a, b and d are polarization-dependent, yet polarization-independent absorbers can be achieved by extending each design into 2D periodic array. Therefore, a 2D periodic plasmonic array can be a structure of choice for designing sensor-on-chip spectroscopic devices with polarization-independent ultra-narrowband absorption.

## 3. Results and Discussions

### 3.1. The Design and Simulation of the Quad-Wavelength IR Sensor

The conceptual design of the Wood's anomaly absorber (WAA) quad-wavelength IR sensor is illustrated in Figure 2a and b. Here, a set of four 2×2 mm$^2$ wavelength-selective pyroelectric sensors with different resonances are arranged into a 1×1 cm$^2$ silicon-based substrate (Figure 2b). Depending on the application, the size of each sensor chip can be scaled down to sub-hundred micrometers. The thermal insulation of each resonant sensor is achieved by a Si$_3$N$_4$ membrane (350-nm-thick) with four slits (20-$\mu$m-width) on the edges of each sensor. In each sensor, a 2D periodic plasmonic disk-on-film array made of gold (Au) works as a narrowband perfect absorber to absorb IR radiation at a resonant wavelength designed under the normal incidence. The absorbed electromagnetic radiation is converted into heat through Joule heating (resistive heat) following the Poynting's theorem described earlier. Hence, the resistive heat in the resonant metal absorber conductively transfers to the ZnO pyroelectric sensor (300 nm thick) to boost electrically polarized charges in the pyroelectric film, generating a temporary voltage at the both ends of the ZnO layer. The temporary voltage is then measured through the top (Au) and bottom (Pt) contacts (100-nm-thick for each film) as the IR sensing signal. The cooling process also gives rise to a reversed temporary voltage signal. The resonance of each sensor relies on the diameter – $d$, the height – $h$ and the periodicity – $p$ of the metallic disk-on-film array.

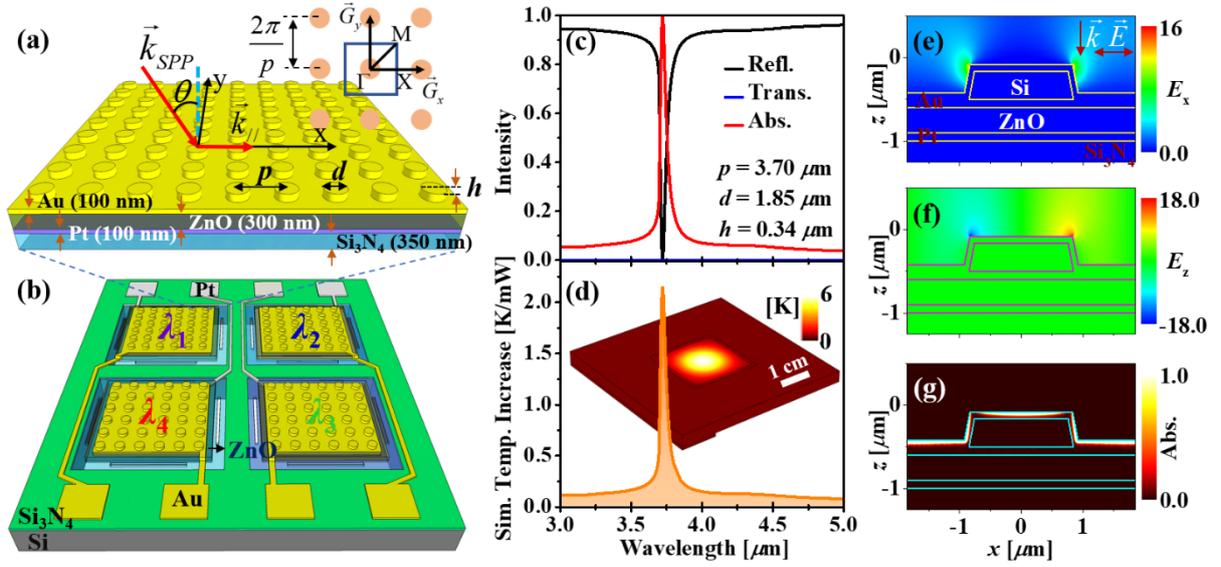

**Figure 2.** (a) Schematic illustration of the WAA IR sensor. (b) Schematic design of the on-chip quad-wavelength membrane pyroelectric sensor. (c) Simulated reflectance, transmittance and absorptivity of a 3.7 $\mu$m periodicity absorber indicates a narrow and perfect absorption. (d) Simulated heat generation spectrum averaged in the ZnO pyroelectric layer of a 3.7 $\mu$m periodicity sensor excited by a 0.5 mm radius gaussian beam with a power of 1 mW. The inset in (d) reveals a 3D heat increase distribution of the sensor. (e) – (g) Simulated distributions of electric fields ($E_x$, $E_z$) and absorption excited at the resonance (3.722 $\mu$m) of a 2D plasmonic array absorber.

The optical resonance of the WAA can be explained through the SPP-photonic coupling in a 2D periodic plasmonic lattice. In a plasmonic square lattice (see the inset in Figure 2a for the reciprocal space), SPPs at the metal-air interface are given by:

$$\left|\vec{k}_{spp}\right| = k_0\sqrt{\frac{\varepsilon_m}{\varepsilon_m+1}}, \tag{2}$$

where $\varepsilon_m$ is the complex permittivity of metal, $k_0 = \frac{2\pi}{\lambda}$. SPPs are excited if their momentums match the momentum of the incident photon and the periodic lattice (momentum conservation):

$$\vec{k}_{spp} = \vec{k}_\| + i\vec{G}_x + j\vec{G}_y, \tag{3}$$

where $\left|\vec{k}_\|\right| = k_0 \sin\theta$ is the projection of momentum of the excitation photon with an incident angle of $\theta$ on the metal surface, $\vec{G}_x$ and $\vec{G}_y$ are two primitive lattice vectors, $i$ and $j$ are integers. In case $\vec{k}_\|$ is oriented along $\vec{G}_x$ ($\Gamma-X$ direction, see the inset in Figure 2a), Equation (3) can be written as:

$$\left|\vec{k}_{spp}\right|^2 = \left|\left(\vec{k}_\| + i\vec{G}_x\right) + j\vec{G}_y\right|^2 = \left(k_0 \sin\theta + iG_x\right)^2 + \left(jG_y\right)^2. \tag{4}$$

From Equations (2) and (4), and with $\left|\vec{G}_x\right| = \left|\vec{G}_y\right| = \frac{2\pi}{p}$, $k_0 = \frac{2\pi}{\lambda}$, the angular dependent dispersion relation of SPPs for a square lattice along $\Gamma-X$ direction is expressed by:

$$\frac{\varepsilon_m}{\varepsilon_m+1} = \sin^2\theta + \left(\frac{2i}{p}\sin\theta\right)\lambda + \left(\frac{i^2+j^2}{p^2}\right)\lambda^2. \tag{5}$$

At normal incidence, the resonant wavelength is simply calculated by: $\lambda = \frac{p}{\sqrt{i^2+j^2}}\sqrt{\frac{\varepsilon_m}{\varepsilon_m+1}}$.

Therefore, at normal incidence, the resonant wavelength of the WAA IR sensor is simply predicted by the periodicity – $p$, although the height and disk size affect the resonance strength. Because of the coupling nature of the plasmonic surface waves with the periodic property, the resonance in the disk-on-film 2D WAA is much narrower compared to the localized plasmon resonance in other absorber structures as proven in Figure 1. From the numerical simulations based on the rigorous coupled-wave analysis (RCWA), we verified that the bandwidth and the absorptivity of the 2D WAA mainly depend on the diameter – $d$ and the height – $h$ of the metallic disk as shown in Figure S1a and b (Supporting Information), respectively. In this quad-wavelength IR sensor, we kept the metallic disks of both four absorbers at the same height and the diameter at a half of the periodicity. Figure 2c shows the simulated optical spectra of a disk-on-film 2D WWA IR sensor having a periodicity of 3.7 $\mu$m, a disk diameter of 1.85 $\mu$m and a disk height of 0.34 $\mu$m under the normal incident excitation. It is worth noting that we adopted a thin Au-covered Si disk array instead of using a submicron-thick Au disk array, which exhibits the same optical properties while saving more Au compared to the identical disk array only made of Au (Figure S2, Supporting Information). As seen from the figure, the plasmonic absorber with 3.7 $\mu$m periodicity exhibits a nearly perfect absorptivity (0.99) resonant peak at 3.722 $\mu$m with a narrow bandwidth of 51 nm ($Q$-factor of 73). With a symmetrical geometry of the 2D plasmonic square lattice adopted here, the WAA does not depend on the polarization (Figure S3a, Supporting Information), which makes the wavelength-selective sensor more applicable in the multifunctional practical applications. The simulated angle-dependent absorptivity of the WAA shown in Figure S3b (Supporting Information) indicates clearly the hybridized nature of SPP waves with the photonic property in a 2D plasmonic square lattice as predicted in Equation (5) (white-dashed curves in Figure S3b).

In order to further elucidate the SPP-photonic coupling in the 2D WAA, electric field distributions in the plasmonic structure were calculated using a full-wave simulation based on the finite-difference time-domain (FDTD) method. Figure 2e and f show the simulated electric field distributions ($E_x$ and $E_z$) of a 3.722 $\mu$m resonant sensor excited at the resonance under normal incidence. It is clearly seen that strong electric nearfields are converged at the edges of the metallic disks. Furthermore, the induced electric field – $E_z$ (Figure 2f) resulted from the coupling between the incident photon and the 2D periodic metallic lattice reveals strongly-coupled nearfields with their phases alternating along the $x$-axis, evidencing that the SPP waves are excited and propagated along the metallic surface. The SPP waves are then strongly damped via metallic losses (Figure 2g) at the Au surface, resulting in a narrow absorption as shown in Figure 2c. This resistive absorption in the metal generates heat due to Joule heating, then heat is conductively transferred to the pyroelectric ZnO layer, producing the temporary voltage signal. The combination of highly selective excitation of grating-coupled surface waves and its swift plasmonic damping is the effective mechanism of the perfect absorption with very high wavelength selectivity. To understand the light-heat conversion and the heat transfer processes in the sensor, we performed multi-physics simulations for the periodic array absorber sensor membrane. Figure 2d and its inset, respectively, present the simulated spectrum of the heat increase averaged in the pyroelectric ZnO layer and the heat distribution of a 3.722 $\mu$m resonant sensor irradiated at the resonance by a 0.5-mm radius gaussian beam with a power of 1 mW. Interestingly, the averaged heat spectrum in the ZnO sensor reveals a resonance at 3.722 $\mu$m, which is exactly the same as the simulated absorption spectrum (Figure 2c). Furthermore, the averaged heat induced by the 2D WAA at the pyroelectric layer increases as high as 2.3 K at thermal equilibrium when it is irradiated at the resonance with a power of 1 mW, which is certainly high enough for the thermal sensor response. Thus, the conceptual design of the quad-wavelength IR sensor is proved and can be adopted for practical applications.

**3.2. Fabrication of MEMS-Based Quad-Wavelength IR Sensor**

To realize the proposed quad-wavelength IR sensor, several steps of direct laser writing lithography involving the film deposition and lift-off, the reactive-ion etching (RIE), and the anisotropic wet-etching were carried out on a 3-inch double side polished Si substrate, where a set of 25 quad-wavelength IR sensors were arranged. The fabrication procedure is depicted in Figure 3, and the details are given in the Experimental Section. As discussed above, instead of using a tall Au disk (340 nm) array for the 2D WAA, an 80-nm-thick Au film coated on a 340-nm-thick Si disks array was adopted to save gold while it retains essentially identical performance to the tall Au disk array. The periodicities of four plasmonic array sensors with their resonances aiming at the transparent atmospheric window in the MWIR region are 3.5 μm, 3.7 μm, 3.8 μm, and 3.9 μm. Here we use a 100 nm thick Pt film deposited by electron beam (EB) evaporation as the bottom electrode, wherein the film also works as an epitaxial substrate with (111) face for growing highly crystalline ZnO(0001) film.[31,45]

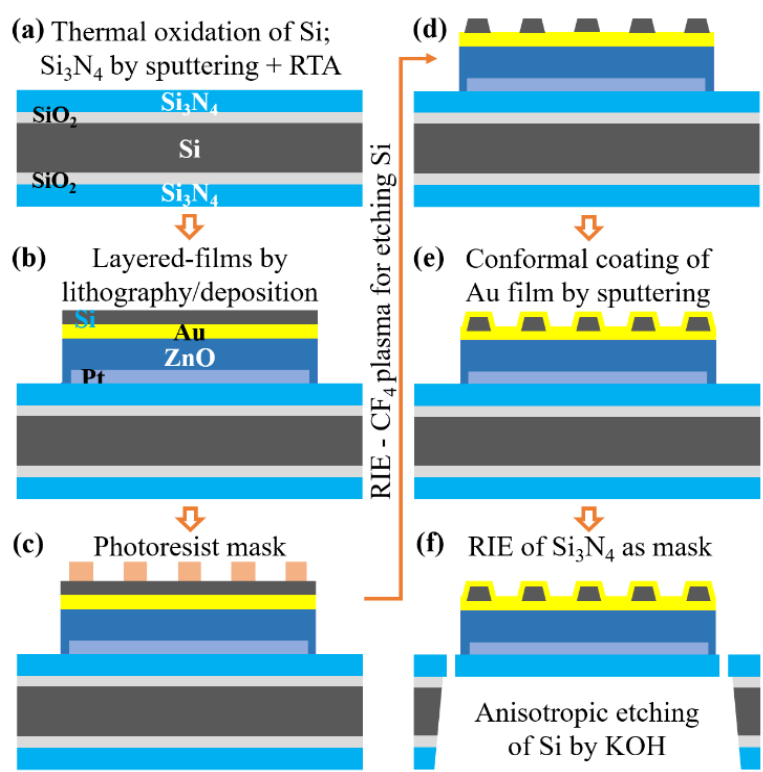

**Figure 3.** Schematic fabrication process of the membrane WAA-IR sensor. (a) Double side polished 3-inch Si wafer with a 100-nm-thermally-oxidized layer and a 350-nm-sputtered Si$_3$N$_4$ film on both sides. (b) Patterning bottom Pt electrode, ZnO pyroelectric film, top Au electrode and Si template film using direct laser writing lithography combined with film deposition and lift-off processes. (c) Patterning photoresist disk array as RIE mask. (d) Patterning Si disk array by RIE with photoresist mask array. (e) Conformal coating of Au layer on Si disk array using sputtering. (f) RIE of top Si$_3$N$_4$ mask for thermal isolation slits, and bottom Si$_3$N$_4$ mask for anisotropic etching of Si, then KOH anisotropic etching of Si for membrane.

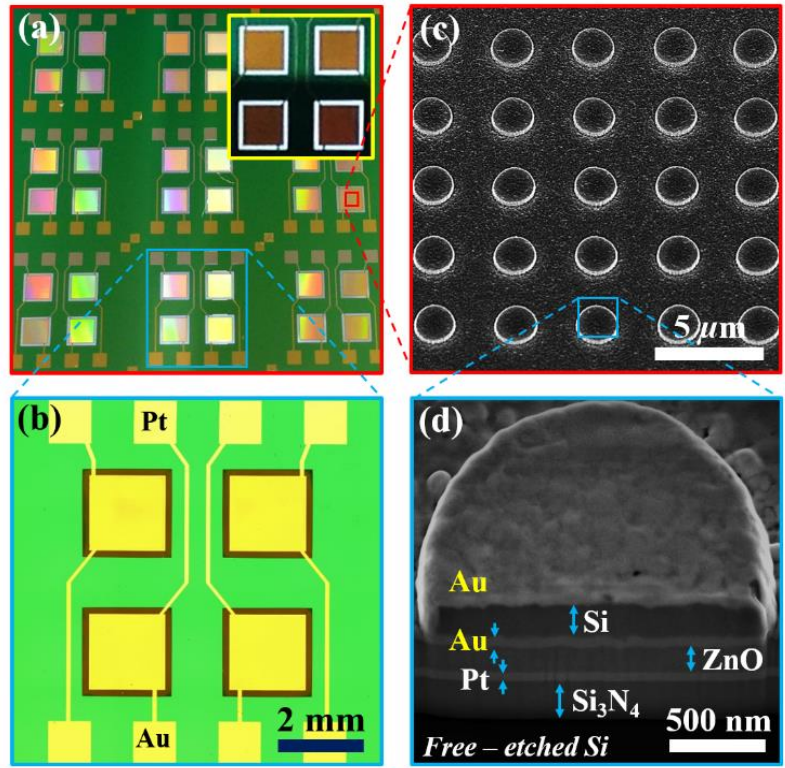

**Figure 4.** (a) A photo of 9 fabricated quad-wavelength membrane pyroelectric sensors. The inset presents a photo of a quad-wavelength membrane sensor where clear transparent white light through the $Si_3N_4$ membrane is visible. (b) Optical microscope image of a quad-wavelength membrane sensor. (c) Top-view SEM image of the fabricated 3.7 $\mu$m periodic resonance sensor. (d) Cross-sectional tilted-view SEM image of a membrane sensor.

Figure 4 summaries the morphological characterizations of the fabricated quad-wavelength IR membrane sensors. Figure 4a is a photo viewed at a set of 9 fabricated quad-wavelength IR sensor chips from a whole 3-inch wafer. The inset in Figure 4a shows a photo of a typical fabricated quad-wavelength sensor taken under the white light illuminated from the bottom, which reveals the optically transparent $Si_3N_4$ membrane layer around each single-wavelength sensor, indicating the fabricated sensors were well suspended from the Si substrate by the $Si_3N_4$ membrane. Figure 4b shows a bright-field optical microscope image scanned over a sensor chip with a scale bar of 2 mm, which approves the dimensional parameters of the fabricated sensor of 2×2 mm$^2$ for each single-wavelength sensor and of 1×1 cm$^2$ for a whole quad-wavelength sensor. A top-view scanning electron microscope (SEM) image of a sensor presented in Figure 4c displays a typical fabricated WAA-IR sensor. A cross-sectional view shown in Figure 4d explores the structural view of the sensor, which clearly evidences that the Au-shell disk supported by the Si-core template was well-constructed using the proposed fabrication process described above. The dimensional parameters of each film layer in the sensor, including the plasmonic disk array, the Au top electrode, the pyroelectric ZnO film, the bottom Pt electrode as well as the $Si_3N_4$ membrane are also clearly seen and verified in Figure 4d, which are the same as designed.

### 3.3. Proof of Concept

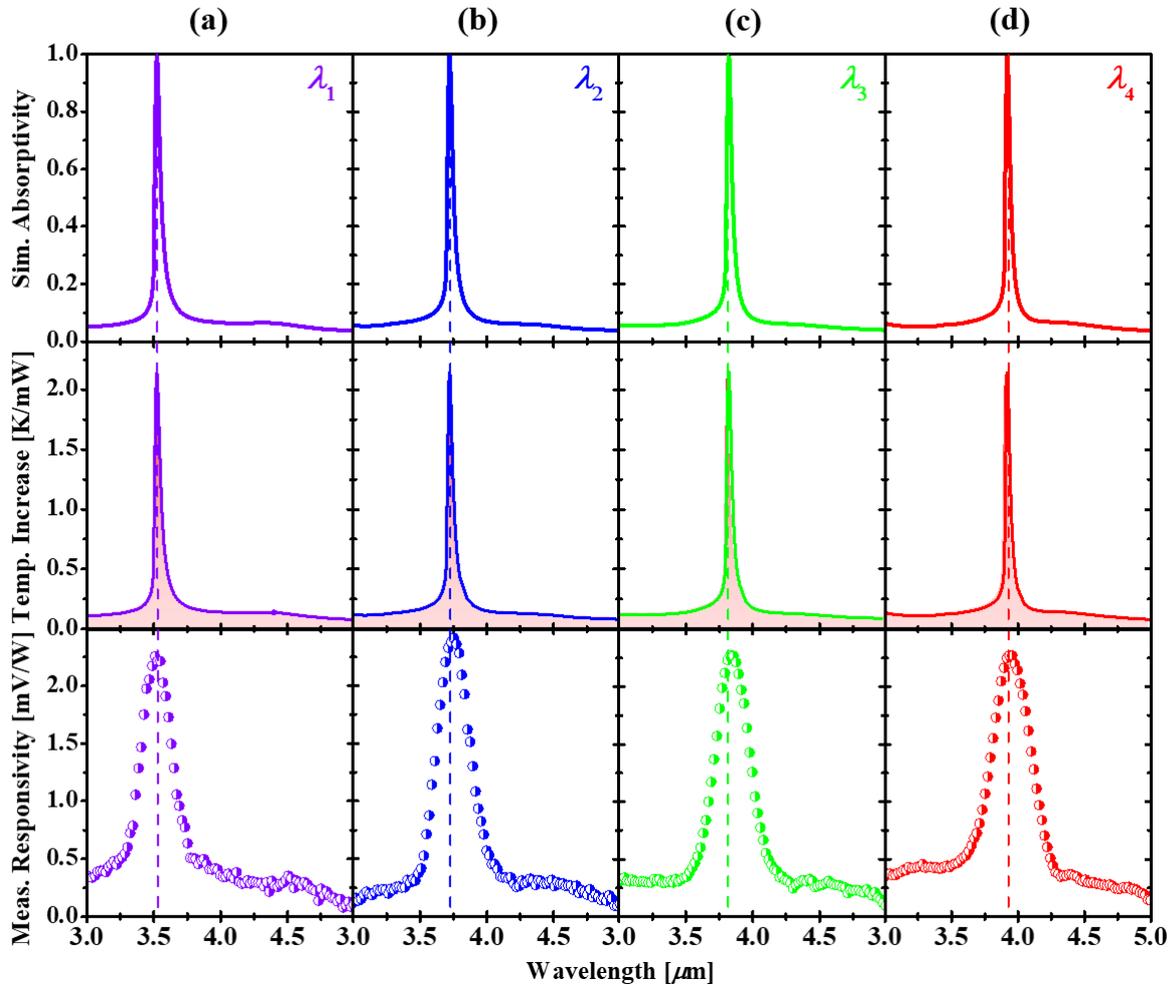

**Figure 5.** (a) – (d) Optical and electrical response of a quad-wavelength membrane pyroelectric sensor with resonances at: (a) 3.522 $\mu$m ($\lambda_1$); (b) 3.722 $\mu$m ($\lambda_2$); (c) 3.822 $\mu$m ($\lambda_3$); (d) 3.922 $\mu$m ($\lambda_4$). From top to bottom panels: Simulated typical absorptivity spectra; simulated heat increase spectra averaged in ZnO layer at thermal equilibrium; and measured spectral response curves of both four wavelengths.

The performance (spectral response) of the fabricated quad-wavelength sensors was characterized using a tunable IR laser system. The measurement setup is detailed in the Experimental Section. Figure 5, from top to bottom, presents the simulated absorptivity spectra, the simulated averaged heat increase spectra and the measured spectral response curves, respectively, of four individual single-wavelength sensors of a typical quad-wavelength sensor. Four single-wavelength sensors were labelled as follows depending on the resonance wavelengths; 3.522 $\mu$m ($\lambda_1$, Figure 5a), 3.722 $\mu$m ($\lambda_2$, Figure 5b), 3.822 $\mu$m ($\lambda_3$, Figure 5c), 3.922 $\mu$m ($\lambda_4$, Figure 5d). Here we used a same disk height – $h$ of 340 nm for all four resonant absorbers, but with different periodicities – $p$ of 3.5 $\mu$m ($\lambda_1$), 3.7 $\mu$m ($\lambda_2$), 3.8 $\mu$m ($\lambda_3$) and 3.9 $\mu$m ($\lambda_4$) wherein each disk diameter was fixed at a half of each periodicity. As seen in Figure 5 (top panels), the simulated responsivity spectra of the quad-wavelength sensor are almost unity (0.99) and narrow (50 nm), which proves that the designed sensor can efficiently absorb IR light at each resonance. Indeed, the simulated temperature increase averaged over the pyroelectric ZnO film at thermal equilibrium taken from both four individual single-wavelength sensors (middle panels in Figure 5) clearly show that the quad-wavelength sensor can efficiently absorb IR light in the narrow spectral bandwidth at the designed resonances, then convert the absorbed resonant IR energy into heat with respect to the absorption spectra, and heat subsequently transfers to the ZnO sensing layer. As expected, the measured spectral response curves shown in the bottom panels of Figure 5 clearly approve the conceptual design of the on-chip multi-wavelength sensor proposed in this work. The four-individual single-wavelength sensors in a quad-wavelength sensor chip exhibit narrow responsivity curves in which their resonances agree well with the simulated absorptivity spectra as pre-designed. The broadening of the measured spectral response curves compared to the simulated absorptivity and temperature increase spectra are due to the broad spectral linewidth ($Q$-factor of 10 – 15) of the IR laser used in the measurement.

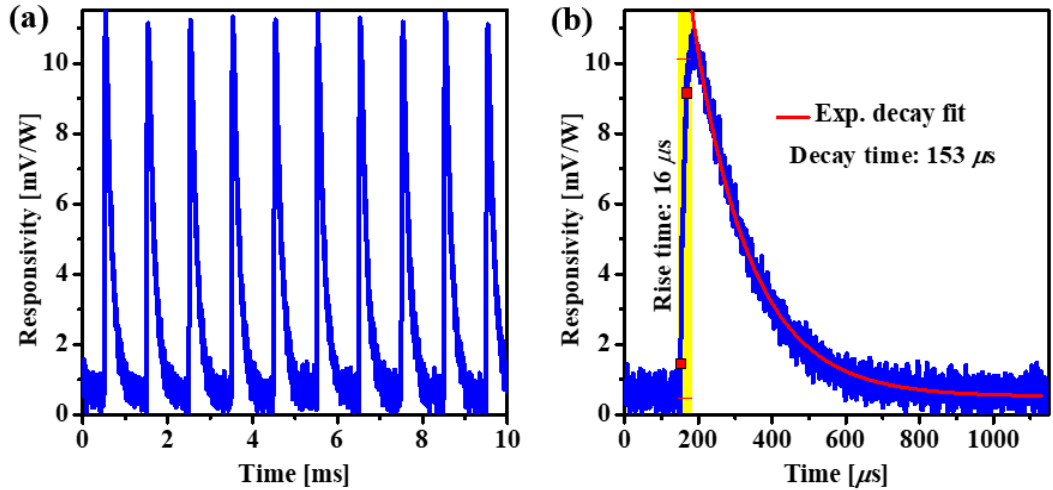

**Figure 6.** Measured temporal response of a 3.722-$\mu$m-resonant sensor chip excited by a 104-fs-pulsed laser at resonance. (a) The stability of the sensor response within 10 pulses duration. (b) The impulse response of the sensor at a single pulse reveals a rise time of 16 $\mu$s and a decay time of 153 $\mu$s.

Since the resonance of the spectroscopic sensor chip was originated from the Wood's anomaly absorption in the 2D periodic plasmonic array, which is angle-dependent, we further performed the angle-dependent responsivity measurement to verify the angular response characteristic of the sensor. Figure S4 (Supporting Information) plots the spectral response curves of a 3.7 $\mu$m periodicity wavelength-selective sensor measured at normal incidence and at 5 degrees and 10 degrees tilted angles. Interestingly, despite the angle-dependent resonance, the responsivity of the sensor at oblique incidence decreases significantly when the incident angle increases, indicating that the fabricated wavelength-selective sensor has high directivity at normal incidence. To understand the dynamic response of the device, the temporal response of a 3.722-$\mu$m-resonant sensor was experimentally measured using a high-performance oscilloscope. As seen in Figure 6, the uniform response of the sensor measured within 10 pulses (Figure 6a) indicates the fast response and the stability of the sensor. The impulse response of the sensor measured with single pulse excitation (Figure 6b) shows a fast response time of 16 $\mu$s and a decay time of 153 $\mu$s, which are applicable to the practical devices.

**3.4. Potential Applications**

With the accessible design and narrowband responsivity at resonances, the sensor-on-chip quad-wavelength IR device developed in this work can be extend into multi-wavelength sensors for practical applications in portable IR spectroscopic devices such as in the true-temperature pyrometry,[46] IR color imaging for environmental detection and materials recognitions,[47] as well as IR remote sensing of atmospheric pollutions for controlling air quality.[48–50] Although the current 2D WAA-based multi-wavelength sensor already shows high directivity at normal incidence, the performance can be further improved by adopting a pinhole aperture, or collimator. Furthermore, if the device is operated with a monochromatic light at its resonance wavelength, it shows an ultra-narrow working angle (below 1 degree, see Figure S5in the Supporting Information), in which one may find angular-sensitive sensing applications such as angular positioning or automotive devices. The current sensor size is on the order of millimeter, however, the sensor can be scaled down to sub-hundred micrometer with the identical design and fabrication, and the ZnO layer can be replaced with more practical thermal sensors such as bolometer or thermopile.

**4. Conclusion**

In conclusion, we have proposed an ultra-narrowband, on-chip multi-wavelength sensor with parallel detection for practical and portable spectroscopic applications. As a proof of concept, we have designed and experimentally demonstrated an on-chip quad-wavelength membrane sensor operating in the MWIR atmospheric window region. The optical and thermal properties of the quad-wavelength membrane sensor were numerically simulated and optimized to have narrow bandwidths ($Q$-factor of 73) and nearly perfect absorptivities (0.99) with an efficient light-heat conversion and a direct conductive heat transfer. The fabricated sensor chips exhibited ultra-sharp spectral response under normal incidence operating in the MWIR region, which appropriately proves that the conceptual design of the WAA multi-wavelength IR sensor is applicable. Our work also provides a clear vison that multi-wavelength spectroscopic device can be fabricated on a single Si chip using CMOS-compatible MEMS design and can be easily extended into multi-color IR devices. With the proposed simple design, the multi-color IR sensors presented in this work exhibit ultra-sharp wavelength resolution of ~50 nm at the normal incidence; and are expected to serve as miniature spectroscopic IR devices useful for true-temperature pyrometry, gas imaging, position and motion sensing with high angular resolution, materials specific imaging, as wells as environmental sensing.

## 5. Experimental Section

*Numerical simulations:* The optical spectra (transmittance, reflectance and absorptivity) of the 2D WAA were simulated using RCWA method (DiffractMOD, Synopsys' RSoft). For the electric field and absorption distributions, a full-wave simulation based on FDTD method (FullWAVE, Synopsys' RSoft) was employed. For FDTD simulation, periodic boundary conditions were applied to the *x*- and *y*-directions, a perfectly matched layer was applied to the *z*-direction, and a mesh size of 2.5 nm was applied for both directions. For both RCWA and FDTD simulations, the excitation electromagnetic field propagated along the –*z*-axis and the electric field oscillated along the *x*-axis, the incident field amplitudes and their phases were normalized to 1. In the electromagnetic simulations, the dielectric functions of Au, Si and $SiO_2$ were taken from the literature[51], the dielectric function of ZnO was taken from literature[31], and $Si_3N_4$ was retrieved from spectroscopic ellipsometry measurement carried out using two ellipsometers (SENTECH, SE 850 DUV for UV – NIR region and SENDIRA for IR region) (see Figure S6, Supporting Information). The heat transfer simulation in the quad-wavelength IR sensor was performed using a finite-element method implemented in a commercial package (COMSOL Multiphysics). The absorptivity of each surface, which was obtained from the RCWA calculation, was pre-defined. The density, thermal conductivity and specific heat capacity of all materials used in the heat transfer simulation are detailed in Table S1 of the Supporting Information.

*Device fabrication:* (See also Figure 3) First, a 3-inch double side polished Si wafer was thermally oxidized at 1150 °C in dry oxidation to form an approximated 100 nm thick $SiO_2$ layer on both sides of the Si wafer. A 350 nm thick $Si_3N_4$ film was then deposited on both sides of the $SiO_2$/Si wafer following a DC (200 W) reactive sputtering recipe with a boron-doped Si target and a mixture of Ar/$N_2$ (18/10 sccm) gases (sputter i-Miller CFS-4EP-LL, Shibaura) (see Figure 2a). Subsequently, a rapid thermal annealing (RTA) process in $N_2$ atmosphere (heating rate of 5 °C per second, keeping constant at 1000 °C for 1 minute, then naturally cooling down) was applied on the sputtered $Si_3N_4$/$SiO_2$/Si substrate to improve the quality (hardness) of the $S_3N_4$ film. Next, we employed a following maskless lithography procedure to create photoresist patterns as mask for the lift-off process of the bottom Pt film electrode. First, double spin-coated AZ-5214E/OFPR-800LB photoresist layers (5000 rpm spin-coating, soft-baking at 90 °C within 5 minutes for each layer) were prepared on the $Si_3N_4$/$SiO_2$/Si substrate. Then, a direct laser writing exposure (405 nm laser wavelength, $\mu$PG 101 Heidelberg Instruments) following a CAD drawing pattern was then applied on the double photoresist layers. After hard baking at 120 °C for 30 second and applying an image reversal exposure under UV light, the exposed double photoresist layers were developed, subsequently rinsed in distilled water and dried using a $N_2$ gas blower. A 100 nm Pt film with a 10 nm adhesive Ti layer for the bottom electrode of the sensor was deposited on the $Si_3N_4$/$SiO_2$/Si substrate with the patterned photoresist mask using an electron beam evaporator (UEP-300-2C, ULVAC). The lift-off process was done using PG remover. For patterning of the pyroelectric ZnO film (by sputtering) and the top Au electrode (by electron beam deposition), the same maskless lithography procedures as the above were applied. It is worth noting that here we used an RF (300 W) sputtering recipe with ZnO target and a mixture of Ar/$O_2$ (16/04 sccm) gases for the epitaxial growth of the highly crystalline ZnO film on Pt bottom film electrode.[30] After preparing the top Au electrode, a 340 nm amorphous Si (boron doped) film as the template layer for Au disk array was patterned on the top Au film electrode (Figure 3b). The photoresist disk arrays (AZ-514E) designed for each quad-wavelength sensor as an RIE mask for etching Si were patterned on the Si template using a direct laser writing lithography process (Figure 3c). Subsequently, an RIE recipe (CH$_4$ plasma, Ulvac CE-300I) was used to etch Si around the photoresist disk mask (Figure 3d). The remaining photoresist was removed by O$_2$ plasma and by acetone. The Au disk array of each 2D periodic plasmonic absorber was finally formed by applying the above maskless lithography procedure with a DC sputtering of 80 nm Au film after a 5 nm adhesive Ti layer (Figure 3e). The quad-wavelength IR sensor chips on the 3-inch wafer were then processed for the thermal isolation with membrane support. Here, AZ-514E photoresist RIE masks (for membrane and for thermal isolation slits around each single-wavelength sensor) of Si$_3$N$_4$ layer was firstly patterned. Then the Si$_3$N$_4$ mask for anisotropic wet-etching of Si was formed using a RIE recipe (CHF$_3$ plasma). After protecting the sensor chips in the top Si wafer by a polymeric protective layer (ProTEK®B3-25 on ProTEK® B3 Primer), the Si substrate at the bottom of each single-wavelength sensor was completely etched by a slow-rate anisotropic wet-etching using a hot KOH solution (8 mg/l, 80 °C). The sensor chip wafer was then kept in PG remover for one day, and finally rinsed by acetone before separating them into 1×1 cm$^2$ quad-wavelength IR membrane sensor chips.

*Characterizations:* SEM images of the fabricated quad-wavelength IR membrane sensor were taken using an SEM (Hitachi SU8230) under an accelerating voltage of 5 kV. For the cross-sectional view image, a focused ion beam miller (Hitachi FB-2100) was used to create a rectangular through hole in a membrane sensor chip. For the spectral response measurement, a tunable IR laser system was used as a tunable excitation source. In this laser system, a one-box ultrafast amplifier system (Solstice, Spectra-Physics) that comprises a mode-locked Ti:sapphire laser and a regenerative amplifier, was used to produce high-power, ultrafast (104 fs), near infrared (800 nm) pulses (1kHz) in a beam of exceptional quality. The spectral range of the output laser from Solstice amplifier system is then extended from UV to IR region and amplified by using a traveling-wave optical parametric amplifier (TOPAS-Prime, Spectra-Physics) combined with a Near-IR, UV, Vis generator (NirUVis, Light Conversion) and a non-collinear difference frequency generator (NDFG, Light Conversion). The output IR laser had following characteristics: board spectral linewidth with the *Q*-factor of about 10 – 15; collimated beam with 1-mm diameter; 1 kHz repetition rate; few milliwatts average power depending on the wavelength. In the measurement, the IR sensors were directly irradiated to the 1-mm-diameter laser. It should be noted that the spectral linewidth of the output IR laser is much broader compared to the absorption bandwidth of IR sensors, which caused the broadening of the spectral response of IR sensors. In the measurement, the signal from quad-wavelength IR membrane sensor excited by the tunable IR laser was firstly pre-amplified using a voltage amplifier (SR560, Stanford Research Systems), and then gained using a lock-in amplifier (LI5640, NF Corporation), and finally measured by a source meter (ADCMT 8252). The spectral power distribution of the output IR laser was measured using a thermal power sensor head (S401C Thorlabs) equipped with a power meter console (PM100D, Thorlabs). The spectral response of each IR sensor was calculated by dividing the spectral output voltage at the IR sensor to the measured spectral power distribution of the IR laser. In the temporal response characteristic of the fabricated IR sensor was measured using a high-performance oscilloscope (500 MHz, Tektronix TDS 520A) combined with SR560 amplifier. The detail of the measurement setup for the spectral response and temporal response of the sensors are illustrated in Figure S7 (Supporting Information).

**Supporting Information**. Further optical properties of the devices and the IR response measurement setup

**Acknowledgements**

This work is partially supported by JSPS KAKENHI (16F16315, JP16H06364, 16H03820), and CREST "Phase Interface Science for Highly Efficient Energy Utilization" (JPMJCR13C3) from Japan Science and Technology Agency. The authors would like to thank all the staffs at Namiki Foundry and Nanofabrication Platform in NIMS for their kind advice and supports, especially Ms. T. Ohki, Ms. T. Sawada and Mr. T. Kishida. Thang. D. Dao would like to thank the fellowship program (P16315) from JSPS.

# On-Chip Quad-Wavelength Pyroelectric Sensor for Spectroscopic Infrared Sensing

Thang Duy Dao[*,1], Satoshi Ishii[1], Anh Tung Doan[1,2], Yoshiki Wada[3], Akihiko Ohi[4], Toshihide Nabatame[4], Tadaaki Nagao[*,1,2]

[1]*International Center for Materials Nanoarchitectonics (MANA), National Institute for Materials Science (NIMS), 1-1 Namiki, Tsukuba, Ibaraki 305-0044, Japan.*
[2]*Department of Condensed Matter Physics, Graduate School of Science, Hokkaido University, Kita 10, Nisi 8, Kita-ku, Sapporo 060-0810, Japan.*
[3]*Research Center for Functional Materials, National Institute for Materials Science (NIMS), 1-1 Namiki, Tsukuba, Ibaraki 305-0044, Japan.*
[4]*Nanotechnology Innovation Station, National Institute for Materials Science (NIMS), 1-1 Namiki, Tsukuba, Ibaraki 305-0044, Japan.*

[*]*Corresponding Authors*
*Thang Duy Dao: Dao.duythang@nims.go.jp; katsiusa@gmail.com*
*Tadaaki Nagao: Nagao.Tadaaki@nims.go.jp*

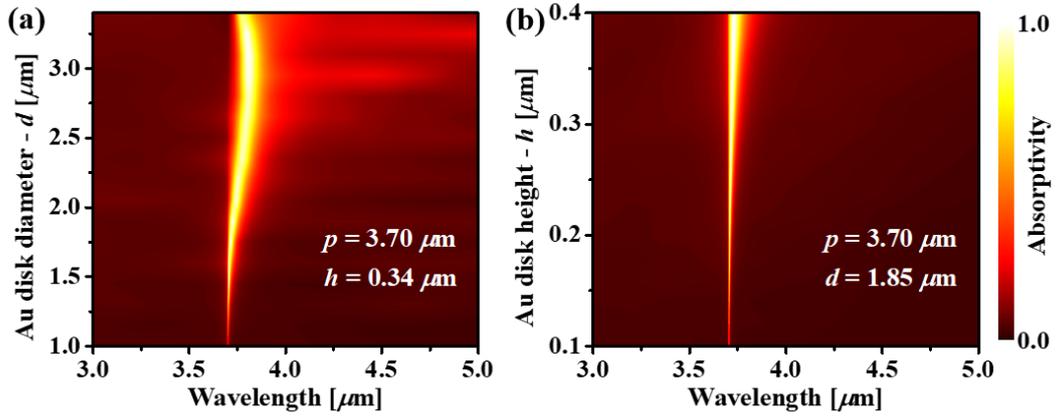

**Figure S1.** Simulated dependences of (a) disk diameter and (b) disk height on the absorptivity of a 3.7 μm single-wavelength membrane pyroelectric sensor.

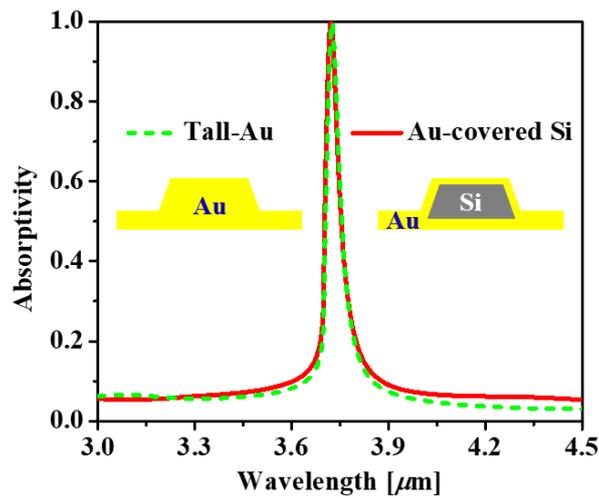

**Figure S2.** Simulated absorptivity spectra of tall Au disk (dashed green curve) and Au-covered Si disk (solid red curve) Wood's anomaly absorbers having identical parameters (periodicity of 3.7 μm, disk diameter of 1.85 μm and height of 0.34 μm). Both two Wood's anomaly absorber configirations exhibit the same performance with a nearly perfect absorptivity (0.99) resonance at 3.722 μm and a narrow bandwidth of 51 nm (*Q*-factor of 73).

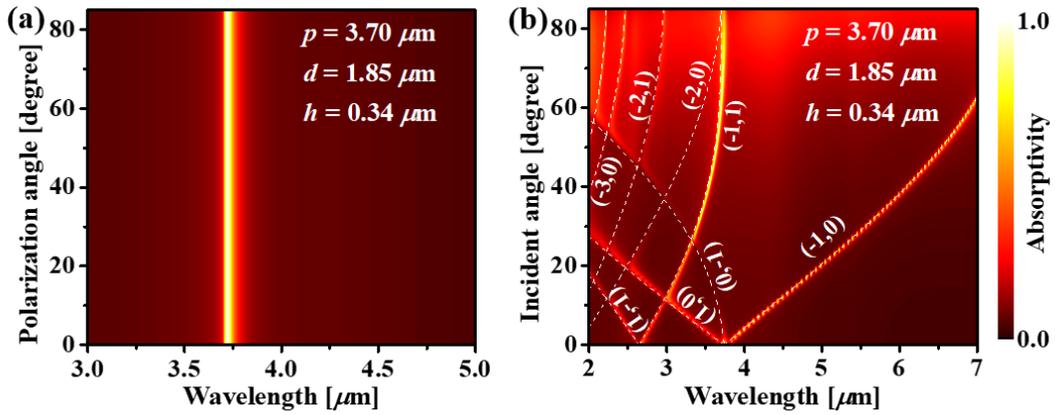

**Figure S3.** Simulated (a) polarization independence and (b) angle dependence on the absorptivity of a 3.7 μm single-wavelength membrane pyroelectric sensor. White-dashed curves indicate SPPs dispersion relation in the 2D periodic plasmonic square lattice.

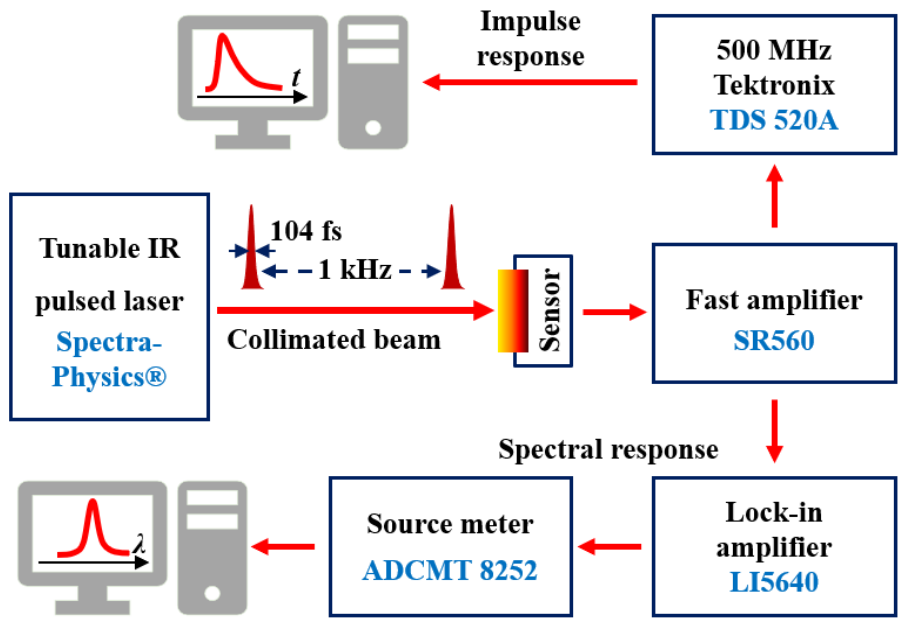

**Figure S4.** Schematic illustration of the measurement setup for the spectral response and temporal response of the on-chip quad-wavelength IR sensor.

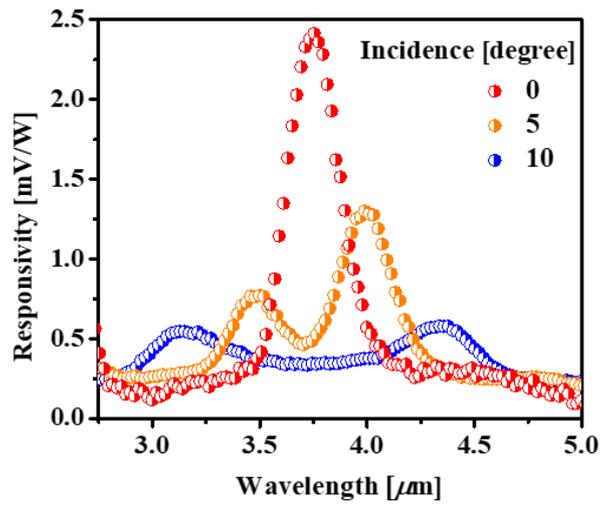

**Figure S5.** Measured angle-dependent spectral response of the 3.722 $\mu$m resonant sensor chip.

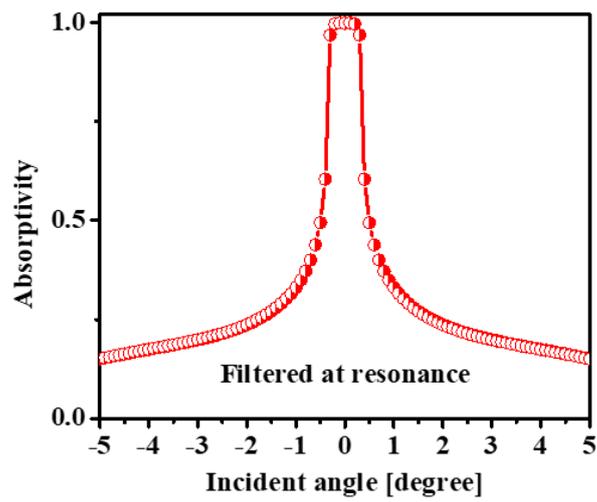

**Figure S6.** Simulated angle-dependent absorptivity of the sensor chip filtered at the resonance.

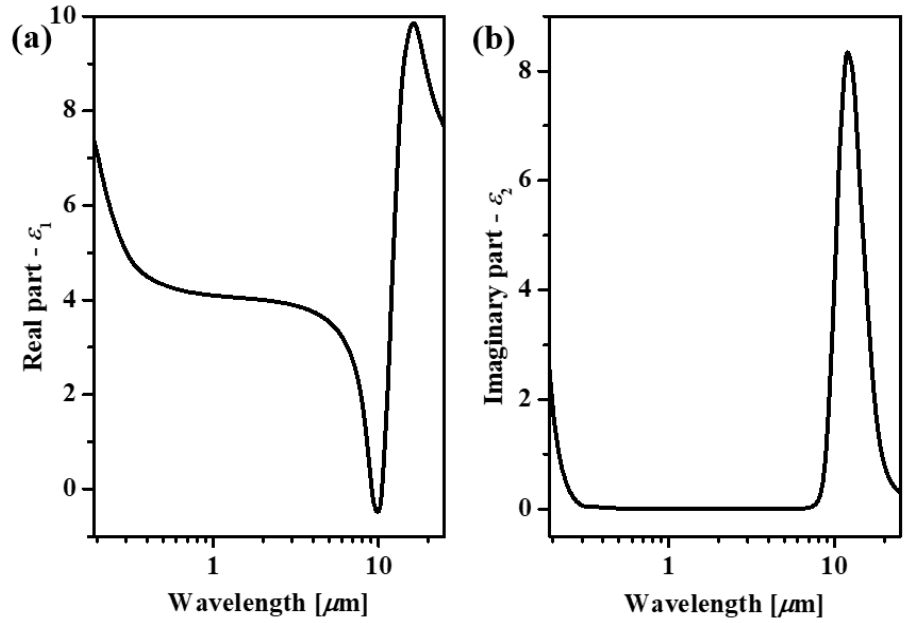

**Figure S7.** (a) Real and (b) imaginary parts of the retrieved complex dielectric function of the sputtered $Si_3N_4$ film.

**Table S1.** The density, thermal conductivity and specific heat capacity of all materials used in the heat transfer simulations.

| Materials | *Density* [g·cm$^{-3}$] | *Thermal conductivity* [W·m$^{-1}$·K$^{-1}$] | *Specific heat capacity* [J·g$^{-1}$·K$^{-1}$] |
|:---:|:---:|:---:|:---:|
| Au | 19.30[S1] | 293[S1] | 0.126[S1] |
| Si | 2.34[S1] | 149[S2] | 0.712[S2] |
| Pt | 21.45[S1] | 71.6[S3] | 0.126[S3] |
| ZnO | 5.675[S4] | 54[S5] | 0.494[S4] |
| $Si_3N_4$ | 3.17[S4] | 43[S4] | 1.1[S4] |
| $SiO_2$ | 2.196[S4] | 1.4[S4] | 0.73[S4] |